\documentclass[showpacs,amsmath,amssymb]{revtex4}
\usepackage{graphicx}
\usepackage{amssymb}

\begin{document}

\title{Chaos in quantum steering in high-dimensional systems}

\author{Guang Ping He}
\email{hegp@mail.sysu.edu.cn}
\affiliation{School of Physics, Sun Yat-sen University, Guangzhou 510275, China}

\begin{abstract}
Quantum steering means that in some bipartite quantum systems the
local measurements on one side can determine the state of the other side.
Here we show that in high-dimensional systems there exists a specific
entangled state which can display a kind of chaos effect when being adopted for steering. That is, a subtle difference in the measurement results on
one side can steer the other side into completely orthogonal states.
Moreover, by expanding the result to infinite-dimensional systems, we find two sets of states for which,
contrary to common belief, even though their density matrices approach
being identical, the steering between them is impossible. This property makes them very useful for quantum cryptography.
\end{abstract}

\maketitle

%%%%%%%%%%%%%%%%%%%%%%%%%%%%%%%%%%%%%%%%%%%%%%%%%%%%%%%%%%%%%%%%%%%%%%%%%%%%%%%%%%%%%%%%%%%%%%%%%%%%%%%%%%%%%%%%%%%%%%%%%%%%%%%%%%%%%%%%%%%%%%%%%%%%%%%%%%%%%%%%%%%%%%%%%%%%%%%%%%%%%%%%%%%%%%%%%%%%%%%%%%%%%%%%%%%%%%%%%%%%%%%%%%%%%%%%%%%%%%%%%%%%%%%%%%%%

\section{Introduction}

The concept of quantum steering (a.k.a., entanglement steering or
Einstein-Podolsky-Rosen steering), first introduced by Schr\"{o}dinger in
1935 \cite{qi1449,qi1450}, is a generalization of the \textquotedblleft
spooky action at a distance\textquotedblright\ proposed by Einstein,
Podolsky and Rosen \cite{qi887}. It means that some bipartite quantum
systems can display a kind of nonlocality, such that the local measurements
on one side of the system can affect (i.e., \textquotedblleft
steer\textquotedblright ) the state of the other side. Recently, with
technological advances in handling quantum entangled states, the theoretical
researches on quantum steering became even more active \cite%
{qi449,qi994,qi1201,qi1229,qi1231,qi1437,qi1438}, with research topics
ranging from the relationship between steering, entanglement, and
nonlocality to the quantifying of steering, and the criterion of
steerability (i.e., whether steering can be observed in a specific quantum
state), etc.

Here we are interested in two questions.

(i) Given a steerable bipartite system $\alpha \otimes \beta $ such that two
local measurements on $\alpha $ can steer $\beta $ into two highly
distinguishable states, respectively, will the two measurements always be
highly distinguishable too?

(ii) Given two sets of states which have identical density matrices, is it
always possible to find a bipartite system that could steer the state of its
subsystem between the elements of the two sets?

In the next section, we will answer the first question by showing that there
exists a specific entangled state in high-dimensional systems, which can
display a kind of chaos effect in steering, i.e., a very subtle difference
in the measurement results on $\alpha $ can steer $\beta $ into completely
orthogonal states. Then in Sec. III, we will answer the second question by
proposing two sets of states in infinite-dimensional systems whose density
matrices approach each other, and show that contrary to common belief, it is
impossible to find a bipartite system that can realize the steering between
these states. The significance of this result will be elaborated in Sec. IV.

\section{Chaos effect in finite-dimensional systems}

Consider a bipartite system $\alpha \otimes \beta $\ prepared in the
entangled state%
\begin{equation}
\left\vert \Omega \right\rangle =\frac{1}{\sqrt{n-1}}\sum_{i=1}^{n-1}\left%
\vert \alpha _{i+}\right\rangle _{\alpha }\left\vert \phi _{i+}\right\rangle
_{\beta }.  \label{omega+}
\end{equation}%
Here $\alpha $ and $\beta $\ are both $n$-dimensional systems, with $%
\left\vert \alpha _{i+}\right\rangle _{\alpha }$ and $\left\vert
i\right\rangle _{\beta }$ ($i=0,...,n-1$) being their orthonormal bases,
respectively (the subscripts $\alpha $ and $\beta $ will be omitted
thereafter), and%
\begin{equation}
\left\vert \phi _{i+}\right\rangle \equiv \frac{1}{\sqrt{2}}(\left\vert
0\right\rangle +\left\vert i\right\rangle )  \label{state+}
\end{equation}%
for $i=1,...,n-1$. Now we will show that Eq. (\ref{omega+}) can display a
kind of chaos effect when we try to steer the state of $\beta $\ by
measuring system $\alpha $.

Defining%
\begin{equation}
\left\vert \phi _{i-}\right\rangle \equiv \frac{1}{\sqrt{2}}(\left\vert
0\right\rangle -\left\vert i\right\rangle )  \label{state-}
\end{equation}%
for $i=1,...,n-1$, and%
\begin{equation}
\left\vert \phi _{n\pm }\right\rangle \equiv \frac{1}{\sqrt{n}}(\left\vert
0\right\rangle \mp \sum\nolimits_{i=1}^{n-1}\left\vert i\right\rangle ),
\label{fai n}
\end{equation}%
we can see that the sets $B_{\pm }\equiv \{\left\vert \phi _{i\pm
}\right\rangle ,i=1,...,n\}$ form two complete (but nonorthogonal) bases of
the $n$-dimensional system $\beta $.

Expanding each $\left\vert \phi _{i+}\right\rangle $\ ($i=1,...,n-1$) in the
basis $B_{-}$, we have%
\begin{equation}
\left\vert \phi _{i+}\right\rangle =\frac{2-n}{n}\left\vert \phi
_{i-}\right\rangle +\sum_{i^{\prime }=1,i^{\prime }\neq i}^{n-1}\frac{2}{n}%
\left\vert \phi _{i^{\prime }-}\right\rangle +\sqrt{\frac{2}{n}}\left\vert
\phi _{n-}\right\rangle .  \label{eq14}
\end{equation}%
Substituting it into Eq. (\ref{omega+}) gives%
\begin{eqnarray}
\left\vert \Omega \right\rangle  &=&\frac{1}{\sqrt{n-1}}(\sum_{i=1}^{n-1}%
\frac{2-n}{n}\left\vert \alpha _{i+}\right\rangle \left\vert \phi
_{i-}\right\rangle +\sum_{i=1}^{n-1}\sum_{i^{\prime }=1,i^{\prime }\neq
i}^{n-1}\frac{2}{n}\left\vert \alpha _{i+}\right\rangle \left\vert \phi
_{i^{\prime }-}\right\rangle )  \nonumber \\
&&+\sqrt{\frac{2}{n}}(\frac{1}{\sqrt{n-1}}\sum_{i=1}^{n-1}\left\vert \alpha
_{i+}\right\rangle )\left\vert \phi _{n-}\right\rangle .
\end{eqnarray}%
Rearranging the second term on the right yields%
\begin{eqnarray}
\left\vert \Omega \right\rangle  &=&\frac{1}{\sqrt{n-1}}\sum_{i=1}^{n-1}(%
\frac{2-n}{n}\left\vert \alpha _{i+}\right\rangle +\frac{2}{n}%
\sum_{i^{\prime }=1,i^{\prime }\neq i}^{n-1}\left\vert \alpha _{i^{\prime
}+}\right\rangle )\left\vert \phi _{i-}\right\rangle   \nonumber \\
&&+\sqrt{\frac{2}{n}}(\frac{1}{\sqrt{n-1}}\sum_{i=1}^{n-1}\left\vert \alpha
_{i+}\right\rangle )\left\vert \phi _{n-}\right\rangle .  \label{omega+'}
\end{eqnarray}%
Define%
\begin{equation}
\left\vert \tilde{\alpha}_{i-}\right\rangle \equiv c_{-}(\frac{2-n}{n}%
\left\vert \alpha _{i+}\right\rangle +\frac{2}{n}\sum_{i^{\prime
}=1,i^{\prime }\neq i}^{n-1}\left\vert \alpha _{i^{\prime }+}\right\rangle )
\label{alfa-}
\end{equation}%
for\ $i=1,...,n-1$, where the normalization constant%
\begin{equation}
c_{-}\equiv \frac{1}{\sqrt{(\frac{2-n}{n})^{2}+(n-2)(\frac{2}{n})^{2}}}=%
\frac{1}{\sqrt{1-4/n^{2}}}.
\end{equation}%
Also define%
\begin{equation}
\left\vert \tilde{\alpha}_{n-}\right\rangle \equiv \frac{1}{\sqrt{n-1}}%
\sum_{i=1}^{n-1}\left\vert \alpha _{i+}\right\rangle ,  \label{alfa n-}
\end{equation}%
then Eq. (\ref{omega+'}) becomes%
\begin{equation}
\left\vert \Omega \right\rangle =\frac{1}{\sqrt{n-1}}\sum_{i=1}^{n-1}\frac{1%
}{c_{-}}\left\vert \tilde{\alpha}_{i-}\right\rangle \left\vert \phi
_{i-}\right\rangle +\sqrt{\frac{2}{n}}\left\vert \tilde{\alpha}%
_{n-}\right\rangle \left\vert \phi _{n-}\right\rangle .  \label{omega-}
\end{equation}

For a given $i$ ($i\in \{1,...,n-1\}$), if we apply the measurement $%
\{\left\vert \tilde{\alpha}_{i-}\right\rangle \left\langle \tilde{\alpha}%
_{i-}\right\vert ,I-\left\vert \tilde{\alpha}_{i-}\right\rangle \left\langle
\tilde{\alpha}_{i-}\right\vert \}$\ (where $I$ is the identity operator) on
system $\alpha $, there is a small but non-vanishing probability that\ the
projection $\left\vert \tilde{\alpha}_{i-}\right\rangle \left\langle \tilde{%
\alpha}_{i-}\right\vert $ will be successful. In this case, Eq. (\ref{omega-}%
) shows that system $\beta $ will collapse to%
\begin{eqnarray}
\left\vert \tilde{\phi}_{i-}\right\rangle  &\equiv &c^{\prime }(\frac{1}{%
c_{-}\sqrt{n-1}}(\left\vert \phi _{i-}\right\rangle +\sum_{i^{\prime
}=1,i^{\prime }\neq i}^{n-1}\left\langle \tilde{\alpha}_{i-}\right.
\left\vert \tilde{\alpha}_{i^{\prime }-}\right\rangle \left\vert \phi
_{i^{\prime }-}\right\rangle )  \nonumber \\
&&+\sqrt{\frac{2}{n}}\left\langle \tilde{\alpha}_{i-}\right. \left\vert
\tilde{\alpha}_{n-}\right\rangle \left\vert \phi _{n-}\right\rangle ),
\label{beta final}
\end{eqnarray}%
where the normalization constant%
\begin{equation}
c^{\prime }=\sqrt{\frac{n(n-1)(n+2)}{(n^{2}+2)}}.
\end{equation}%
From Eqs. (\ref{alfa-}) and (\ref{alfa n-}) we can derive%
\begin{equation}
\left\langle \tilde{\alpha}_{i-}\right. \left\vert \tilde{\alpha}_{i^{\prime
}-}\right\rangle =\frac{-4}{n^{2}-4}  \label{ortho}
\end{equation}%
for any $i\neq i^{\prime }$\ ($i,i^{\prime }\in \{1,...,n-1\}$), and%
\begin{equation}
\left\langle \tilde{\alpha}_{i-}\right. \left\vert \tilde{\alpha}%
_{n-}\right\rangle =\frac{\sqrt{n-2}}{\sqrt{n-1}\sqrt{n+2}}
\end{equation}%
for any $i$\ ($i\in \{1,...,n-1\}$). Substituting them and $c_{-}=1/\sqrt{%
1-4/n^{2}}$ into Eq. (\ref{beta final}), we have%
\begin{eqnarray}
\left\vert \tilde{\phi}_{i-}\right\rangle  &=&c^{\prime }(\frac{\sqrt{%
1-4/n^{2}}}{\sqrt{n-1}}(\left\vert \phi _{i-}\right\rangle -\frac{4}{n^{2}-4}%
\sum_{i^{\prime }=1,i^{\prime }\neq i}^{n-1}\left\vert \phi _{i^{\prime
}-}\right\rangle )  \nonumber \\
&&+\frac{\sqrt{2/n}\sqrt{n-2}}{\sqrt{n-1}\sqrt{n+2}}\left\vert \phi
_{n-}\right\rangle ).  \label{beta final2}
\end{eqnarray}%
Thus we obtain Result 1.

\bigskip

\textbf{Result 1}: \textbf{If we manage to project system }$\alpha $\textbf{%
\ into the state }$\left\vert \tilde{\alpha}_{i-}\right\rangle $\textbf{\
defined in Eq. (\ref{alfa-}), then }$\beta $\textbf{\ will collapse into the
state }$\left\vert \tilde{\phi}_{i-}\right\rangle $\textbf{\ in Eq. (\ref%
{beta final2}).}

\bigskip

On the other hand, from Eq. (\ref{omega+}) we can see that for the same
given $i$, if we apply the measurement $\{\left\vert \alpha
_{i+}\right\rangle \left\langle \alpha _{i+}\right\vert ,I-\left\vert \alpha
_{i+}\right\rangle \left\langle \alpha _{i+}\right\vert \}$\ on system $%
\alpha $ instead, there is also a small but non-vanishing probability that\
the projection $\left\vert \alpha _{i+}\right\rangle \left\langle \alpha
_{i+}\right\vert $ will be successful. In this case, the resultant state of
system $\beta $\ can be obtained directly from Eq. (\ref{omega+}), yielding
Result 2.

\bigskip

\textbf{Result 2}: \textbf{If we manage to project system }$\alpha $\textbf{%
\ into the state }$\left\vert \alpha _{i+}\right\rangle $\textbf{, then }$%
\beta $\textbf{\ will collapse into the state }$\left\vert \phi
_{i+}\right\rangle $\textbf{\ in Eq. (\ref{state+}).}

\bigskip

Now let us study the relationship between the states in Results 1 and 2.
From Eq. (\ref{alfa-}) we find%
\begin{equation}
|\left\langle \alpha _{i+}\right. \left\vert \tilde{\alpha}%
_{i-}\right\rangle |^{2}=|\frac{2-n}{n}c_{-}|^{2}=1-\frac{4}{n+2},
\label{old eq31}
\end{equation}%
i.e., $\left\vert \alpha _{i+}\right\rangle $ and $\left\vert \tilde{\alpha}%
_{i-}\right\rangle $ are very close to each other when $n$ is high. In
contrast, multiplying $\left\langle \phi _{i+}\right\vert $\ by the
right-hand side of Eq. (\ref{beta final2}), we have%
\begin{eqnarray}
\left\langle \phi _{i+}\right. \left\vert \tilde{\phi}_{i-}\right\rangle
&=&c^{\prime }(\frac{\sqrt{1-4/n^{2}}}{\sqrt{n-1}}(\left\langle \phi
_{i+}\right. \left\vert \phi _{i-}\right\rangle -\frac{4}{n^{2}-4}%
\sum_{i^{\prime }=1,i^{\prime }\neq i}^{n-1}\left\langle \phi _{i+}\right.
\left\vert \phi _{i^{\prime }-}\right\rangle )  \nonumber \\
&&+\frac{\sqrt{2/n}\sqrt{n-2}}{\sqrt{n-1}\sqrt{n+2}}\left\langle \phi
_{i+}\right. \left\vert \phi _{n-}\right\rangle )  \nonumber \\
&=&c^{\prime }(\frac{\sqrt{1-4/n^{2}}}{\sqrt{n-1}}(0-\frac{4}{n^{2}-4}(n-2)%
\frac{1}{2})  \nonumber \\
&&+\frac{\sqrt{2/n}\sqrt{n-2}}{\sqrt{n-1}\sqrt{n+2}}\sqrt{2/n})  \nonumber \\
&=&0
\end{eqnarray}%
for any $n$, i.e., $\left\vert \phi _{i+}\right\rangle $ and $\left\vert
\tilde{\phi}_{i-}\right\rangle $ are always strictly orthogonal to each
other. Recall that Results 1 and 2 mean that projecting system $\alpha $
into $\left\vert \alpha _{i+}\right\rangle $ (or $\left\vert \tilde{\alpha}%
_{i-}\right\rangle $) will make system $\beta $ collapse into $\left\vert
\phi _{i+}\right\rangle $ (or $\left\vert \tilde{\phi}_{i-}\right\rangle $).
Then with the fact that $\left\vert \alpha _{i+}\right\rangle $ and $%
\left\vert \tilde{\alpha}_{i-}\right\rangle $ are very close while $%
\left\vert \phi _{i+}\right\rangle $ and $\left\vert \tilde{\phi}%
_{i-}\right\rangle $ are strictly orthogonal, we obtain Conclusion 1.

\bigskip

\textbf{Conclusion 1:} There exists a certain form of states [see, e.g., Eq.
(\ref{omega+})] in high-dimensional systems, such that when it is adopted
for quantum steering, we can observe the following \textquotedblleft
chaos\textquotedblright\ behavior: \textbf{a very subtle difference on the
measurement results on }$\alpha $\textbf{\ can lead to completely orthogonal
steering results on }$\beta $\textbf{.}

\bigskip

However, at first glance this chaos effect seems hard to find because, as
stated above, for a given $i$ if we apply the measurement $\{\left\vert
\tilde{\alpha}_{i-}\right\rangle \left\langle \tilde{\alpha}_{i-}\right\vert
,I-\left\vert \tilde{\alpha}_{i-}\right\rangle \left\langle \tilde{\alpha}%
_{i-}\right\vert \}$ on system $\alpha $, the projection $\left\vert \tilde{%
\alpha}_{i-}\right\rangle \left\langle \tilde{\alpha}_{i-}\right\vert $ will
be successful with a very small probability only. Meanwhile, for any finite $%
n$, Eqs. (\ref{alfa-}) and (\ref{alfa n-}) show that $\{\left\vert \tilde{%
\alpha}_{i-}\right\rangle ,i=1,...,n\}$ is a nonorthogonal set, so that we
cannot use it as an orthogonal measurement basis. Nevertheless, $%
\{\left\vert \alpha _{i+}\right\rangle ,i=0,1,...,n-1\}$ is an orthogonal
measurement basis, as this is how it was defined below Eq. (\ref{omega+}).
Suppose that we prepare the state $\left\vert \Omega \right\rangle $ in Eq. (%
\ref{omega+}) and apply the complete measurement $\{\left\vert \alpha
_{i+}\right\rangle \left\langle \alpha _{i+}\right\vert ,i=0,1,...,n-1\}$\
on $\alpha $. In this case the measurement device will surely produce an
output $i_{0}\in \{1,...,n-1\}$. Then from Result 2, we will assume that $%
\alpha $ was projected to $\left\vert \alpha _{i_{0}+}\right\rangle $
successfully and system $\beta $ was steered into $\left\vert \phi
_{i_{0}+}\right\rangle $. But in practice, there could be a chance that, due
to the imprecision of the measurement device, $\alpha $ may actually be
projected into $\left\vert \tilde{\alpha}_{i_{0}-}\right\rangle $ since Eq. (%
\ref{old eq31}) indicates that it is very close to $\left\vert \alpha
_{i_{0}+}\right\rangle $ when $n$ is high. Theoretically, this is equivalent
to applying the measurements $\{\left\vert \tilde{\alpha}_{i_{0}-}\right%
\rangle \left\langle \tilde{\alpha}_{i_{0}-}\right\vert ,I-\left\vert \tilde{%
\alpha}_{i_{0}-}\right\rangle \left\langle \tilde{\alpha}_{i_{0}-}\right%
\vert \}$, and the projection $\left\vert \tilde{\alpha}_{i_{0}-}\right%
\rangle \left\langle \tilde{\alpha}_{i_{0}-}\right\vert $\ was successful.
Thus $\beta $ was actually steered into $\left\vert \tilde{\phi}%
_{i_{0}-}\right\rangle $, i.e., the chaos effect can indeed occur physically.

It is also worth further studying whether such chaos could be at least one
of the origins of quantum uncertainty in measurements. We would like to
leave it open for future discussions.

\section{Anomalous behavior in infinite-dimensional systems}

When taking the $n\rightarrow \infty $\ limit in the above equations, the
result will be even more interesting. We shall prove below that it will lead
to Conclusion 2.

\bigskip

\textbf{Conclusion 2:} \textbf{In infinite-dimensional systems, it is
\textit{impossible} to construct a bipartite system }$\alpha \otimes \beta $%
\textbf{, such that by the local measurement on system }$\alpha $\textbf{\
alone, the state of system }$\beta $\textbf{\ can be steered between the
following two sets of evenly distributed states}%
\begin{equation}
\{\left\vert \phi _{i+}\right\rangle \equiv \frac{1}{\sqrt{2}}(\left\vert
0\right\rangle +\left\vert i\right\rangle ),i=1,...,n-1\}  \label{set+}
\end{equation}%
\textbf{and}%
\begin{equation}
\{\left\vert \phi _{i-}\right\rangle \equiv \frac{1}{\sqrt{2}}(\left\vert
0\right\rangle -\left\vert i\right\rangle ),i=1,...,n-1\},  \label{set-}
\end{equation}%
\textbf{where }$n\rightarrow \infty $\textbf{.}

\bigskip

That is, we are going to prove that for any bipartite system $\alpha \otimes
\beta $, it is impossible to find two \textit{different} measurements $M_{+}$
and $M_{-}$, such that applying $M_{+}$\ on $\alpha $\ will make $\beta $\
collapse to one of the states in set $\{\left\vert \phi _{i+}\right\rangle
\} $, while applying $M_{-}$\ on $\alpha $\ instead will make $\beta $\
collapse to one of the states in set $\{\left\vert \phi _{i-}\right\rangle
\} $.

%Our proof makes use of the method \textquotedblleft reduction to
%absurdity\textquotedblright . That is, we shall show that if such a
%bipartite system $\alpha \otimes \beta $\ can be constructed, then it will
%lead to an absurd result.

Let us start the proof by assuming that there is a bipartite system $\alpha
\otimes \beta $ which can steer the state of $\beta $ to the elements of set
$\{\left\vert \phi _{i+}\right\rangle \}$. Then there must exist a local
measurement $M_{+}$, such that applying it on $\alpha $ will yield an index $%
i$ of state $\left\vert \phi _{i+}\right\rangle $ to which $\beta $ will
have collapsed. Denote the eigenstates of $M_{+}$ as $\left\vert \alpha
_{i+}\right\rangle $ ($i=0,...,n-1$), i.e.,%
\begin{equation}
M_{+}=\sum_{i=0}^{n-1}i\left\vert \alpha _{i+}\right\rangle \left\langle
\alpha _{i+}\right\vert .  \label{M+}
\end{equation}%
Then the state of $\alpha \otimes \beta $\ can surely be written in the form
of Eq. (\ref{omega+}). In brief, \textbf{any system that can steer the state
of its part into set }$\{\left\vert \phi _{i+}\right\rangle \}$\textbf{\
will take the form of Eq. (\ref{omega+}) as long as the basis }$\{\left\vert
\alpha _{i+}\right\rangle ,i=0,...,n-1\}$\textbf{\ is properly defined.}

Now the question is whether such a system can be steered into the states in
set $\{\left\vert\phi _{i-}\right\rangle \}$. If the answer is yes, then it
means that there exists another local measurement $M_{-}$, such that
applying it on $\alpha $ will yield an index $i$ of state $\left\vert \phi
_{i-}\right\rangle $ to which $\beta $ will have collapsed. Denote the
eigenstates of $M_{-}$ as $\left\vert \alpha _{i-}\right\rangle $ ($%
i=0,...,n-1$), i.e.,%
\begin{equation}
M_{-}=\sum_{i=0}^{n-1}i\left\vert \alpha _{i-}\right\rangle \left\langle
\alpha _{i-}\right\vert .  \label{M-}
\end{equation}%
Then the same $\left\vert \Omega \right\rangle $ in Eq. (\ref{omega+})
should also be able to be expanded in the basis $\{\left\vert \alpha
_{i-}\right\rangle \}$ as%
\begin{equation}
\left\vert \Omega \right\rangle =\frac{1}{\sqrt{n-1}}\sum_{i=1}^{n-1}\left%
\vert \alpha _{i-}\right\rangle \left\vert \phi _{i-}\right\rangle .
\label{omega-2}
\end{equation}%
Let us study \textit{what is the relationship between the measurements }$%
M_{+}$\textit{\ and }$M_{-}$\textit{.}

Since the equations in the previous section are valid for any $n$, they also
apply when $n\rightarrow \infty $. So we can still obtain Results 1 and 2
that projecting system $\alpha $ into $\left\vert \alpha _{i+}\right\rangle $
(or $\left\vert \tilde{\alpha}_{i-}\right\rangle $) will make system $\beta $
collapse into $\left\vert \phi _{i+}\right\rangle $ (or $\left\vert \tilde{%
\phi}_{i-}\right\rangle $).

But multiplying $\left\langle \alpha _{i+}\right\vert $\ by the right-hand
side of Eq. (\ref{alfa-}), we have%
\begin{equation}
\left\langle \alpha _{i+}\right. \left\vert \tilde{\alpha}_{i-}\right\rangle
=\frac{2-n}{n}c_{-}=-\sqrt{1-\frac{4}{n+2}}.  \label{inner}
\end{equation}%
Likewise, multiplying $\left\langle \phi _{i-}\right\vert $\ by the
right-hand side of Eq. (\ref{beta final2}) gives%
\begin{equation}
\left\langle \phi _{i-}\right. \left\vert \tilde{\phi}_{i-}\right\rangle =%
\sqrt{1-\frac{2n+2}{n^{2}+2}}.  \label{new eq25}
\end{equation}%
Therefore, in the $n\rightarrow \infty $ limit we have%
\begin{equation}
\left\vert \tilde{\alpha}_{i-}\right\rangle =-\left\vert \alpha
_{i+}\right\rangle   \label{alfa- infinite}
\end{equation}%
and%
\begin{equation}
\left\vert \tilde{\phi}_{i-}\right\rangle =\left\vert \phi
_{i-}\right\rangle .  \label{old eq26}
\end{equation}%
That is, in infinite-dimensional systems, Result 2 becomes Result 2'.

\bigskip

\textbf{Result 2': If we manage to project system }$\alpha $\textbf{\ into }$%
-\left\vert \alpha _{i+}\right\rangle $\textbf{, then }$\beta $\textbf{\
will collapse into }$\left\vert \phi _{i-}\right\rangle $\textbf{.}

\bigskip

Now recall that we assumed that Eq. (\ref{omega-2}) also applies to this
system, which implies that if we manage to project system $\alpha $\ into $%
\left\vert \alpha _{i-}\right\rangle $, then $\beta $\ will collapse into $%
\left\vert \phi _{i-}\right\rangle $. Comparing with Result 2' and Eq. (\ref%
{alfa- infinite}), we know that%
\begin{equation}
\left\vert \alpha _{i-}\right\rangle =\left\vert \tilde{\alpha}%
_{i-}\right\rangle =-\left\vert \alpha _{i+}\right\rangle
\label{alfa result}
\end{equation}%
and therefore%
\begin{equation}
\left\vert \alpha _{i-}\right\rangle \left\langle \alpha _{i-}\right\vert
=\left\vert \alpha _{i+}\right\rangle \left\langle \alpha _{i+}\right\vert
\end{equation}%
for $i=1,...,n-1$. Also, since $\sum_{i=0}^{n-1}\left\vert \alpha _{i\pm
}\right\rangle \left\langle \alpha _{i\pm }\right\vert =I$ with $I$ being
the identity matrix, we have%
\begin{equation}
\left\vert \alpha _{0-}\right\rangle \left\langle \alpha _{0-}\right\vert
=I-\sum_{i=1}^{n-1}\left\vert \alpha _{i-}\right\rangle \left\langle \alpha
_{i-}\right\vert =I-\sum_{i=1}^{n-1}\left\vert \alpha _{i+}\right\rangle
\left\langle \alpha _{i+}\right\vert =\left\vert \alpha _{0+}\right\rangle
\left\langle \alpha _{0+}\right\vert .
\end{equation}%
Substituting these into Eqs. (\ref{M+}) and (\ref{M-}), we find%
\begin{equation}
M_{+}=M_{-}.
\end{equation}%
These equations mean that if we want to collapse $\beta $ into one of the
states in the set $\{\left\vert \phi _{i+}\right\rangle \}$ (or $%
\{\left\vert \phi _{i-}\right\rangle \}$), then we should measure $\alpha $
in the basis $\{\left\vert \alpha _{i+}\right\rangle ,i=0,...,n-1\}$\ (or $%
\{\left\vert \alpha _{i-}\right\rangle =-\left\vert \alpha
_{i+}\right\rangle ,i=0,...,n-1\}$). However, as the global negative sign
before the state vector has no physical meaning, these two bases are
actually the same. Consequently, the measurements $M_{+}$ and $M_{-}$ on $%
\alpha $ for collapsing $\beta $ to an element of the sets $\{\left\vert
\phi _{i+}\right\rangle \}$ and $\{\left\vert \phi _{i-}\right\rangle \}$,
respectively, are no longer two different measurements when $n\rightarrow
\infty $.

Thus it is shown that \textbf{for any bipartite system }$\alpha \otimes
\beta $\textbf{\ which can steer the state of }$\beta $\textbf{\ to the
elements of set }$\{\left\vert \phi _{i+}\right\rangle \}$\textbf{, if we
want to steer the state of }$\beta $\textbf{\ to the elements of set }$%
\{\left\vert \phi _{i-}\right\rangle \}$\textbf{\ instead, we will find that
the corresponding measurement }$M_{-}$\textbf{\ is completely
indistinguishable from the measurement }$M_{+}$\textbf{\ for steering }$%
\beta $\textbf{\ to }$\{\left\vert \phi _{i+}\right\rangle \}$\textbf{.} %%
%Therefore, if we assume that Eqs. (\ref{omega+}) and (\ref{omega-2}) can be %valid simultaneously for the same bipartite system, then it will lead to the %absurd result that the same measurement result $\left\vert \alpha %_{i+}\right\rangle =-\left\vert \alpha
%_{i-}\right\rangle $ of $\alpha $ can collapse $\beta $ to both $%
%\left\vert \phi _{i+}\right\rangle $ and $\left\vert \phi
%_{i-}\right\rangle $, while $%
%\left\vert \phi _{i+}\right\rangle $ is obviously orthogonal to $\left\vert %\phi_{i-}\right\rangle $.
This completes the proof that it is impossible to find a bipartite system
which can steer one of its part between the states $\{\left\vert \phi
_{i+}\right\rangle \}$ and $\{\left\vert \phi _{i-}\right\rangle \}$ by
choosing different measurements on the other part.

Some might wonder whether it is legitimate to take the $n\rightarrow \infty $
limit in our above equations and those in the Appendix. But in fact this
impossibility of steering can be found even for finite $n$, as long as it is
sufficiently high. This is because all physical measurement devices are
subjected to the uncertainty principle, so that they cannot be manufactured
and adjusted with unlimited precision. When $n$ rises to an extremely high
(but still finite) value, even though Eq. (\ref{alfa- infinite}) is not
rigorously satisfied, $\left\langle \alpha _{i+}\right. \left\vert \tilde{%
\alpha}_{i-}\right\rangle $ will become so close to $-1$ [as shown by Eq. (%
\ref{inner})] that distinguishing $\left\vert \tilde{\alpha}%
_{i-}\right\rangle $ and $-\left\vert \alpha _{i+}\right\rangle $ will
require extremely subtle adjustment on the measurement devices (such as
controlling the width of the slits in interference systems, twisting the
angles of the lens, etc.), which falls within the Planck scale. Such subtle
adjustment is inaccessible, both theoretically and practically. Thus $%
\left\vert \tilde{\alpha}_{i-}\right\rangle $ and $-\left\vert \alpha
_{i+}\right\rangle $ become physically indistinguishable, so that even if
the measurements $M_{+}$ and $M_{-}$ are not strictly equal to each other,
distinguishing them is still impossible. As a consequence, the local
measurements on $\alpha $ are insufficient for steering $\beta $ between $%
\{\left\vert \phi _{i+}\right\rangle \}$ and $\{\left\vert \phi
_{i-}\right\rangle \}$\ even in such a finite-dimensional bipartite system.

\section{Significance of the result}

It might look trivial just to find two sets of states, and prove that it is
impossible to realize the quantum steering between them. But a very
important feature of our result, is that the two sets of states $%
\{\left\vert \phi _{i+}\right\rangle \}$ and $\{\left\vert \phi
_{i-}\right\rangle \}$ also display another distinct property. As shown in
the Appendix, the trace distance between the density matrices $\rho _{+}$\
and $\rho _{-}$\ corresponding to $\{\left\vert \phi _{i+}\right\rangle \}$
and $\{\left\vert \phi _{i-}\right\rangle \}$, respectively, is $D(\rho
_{+},\rho _{-})=1/\sqrt{n-1}$, which drops as $n$ increases. Therefore, when
$n$ is extremely high, $\{\left\vert \phi _{i+}\right\rangle \}$ and $%
\{\left\vert \phi _{i-}\right\rangle \}$ will become physically
indistinguishable.

This property is important because in quantum steering, there is a
well-known result called the Hughston-Jozsa-Wootters (HJW) theorem \cite%
{qi73}. It also appears under different names (e.g., the Uhlmann theorem)
and presentations in literature \cite{qbc8,qi1454,qi1455}. A very concise
summary of its conclusion can be found in Ref. \cite{qi56}, which goes as
follows.

\bigskip

\textit{Let }$\psi _{1}$\textit{,\ }$\psi _{2}$\textit{,\ ..., }$\psi _{m}$%
\textit{\ and }$\psi _{1}^{\prime }$\textit{,\ }$\psi _{2}^{\prime }$\textit{%
,\ ..., }$\psi _{m^{\prime }}^{\prime }$\textit{\ be two sets of possible
quantum states with associated probabilities described by an identical
density matrix }$\rho $\textit{. It is possible to construct a composite
system }$\alpha \otimes \beta $\textit{\ such that }$\beta $\textit{\ alone
has density matrix }$\rho $\textit{\ and such that there exists a pair of
measurements }$M_{\psi }$\textit{\ and }$M_{\psi ^{\prime }}$\textit{\ with
the property that applying }$M_{\psi }$\textit{\ (}$M_{\psi ^{\prime }}$%
\textit{) to }$\alpha $\textit{\ yields an index }$i$\textit{\ of state }$%
\psi _{i}$\textit{\ (}$\psi _{i}^{\prime }$\textit{) to which }$\beta $%
\textit{\ will have collapsed.}

\bigskip

The original proofs % \cite{qi73,qbc8,qi1454,qi1455}
of the HJW theorem are defined only for finite-dimensional ensembles. Later,
they were generalized to the infinite-dimensional ensembles \cite%
{qbc35,qi1473}. But our result seems to conflict with this theorem, as we
showed that for extremely high $n$, constructing such a composite system to
steer between the two sets $\{\left\vert \phi _{i+}\right\rangle \}$ and $%
\{\left\vert \phi _{i-}\right\rangle \}$ defined in Eqs. (\ref{set+}) and (%
\ref{set-}) is impossible, even though their trace distance $D(\rho
_{+},\rho _{-})$ can be made arbitrarily small.

Nevertheless, after checking the existing proofs of the HJW theorem \cite%
{qi73,qbc8,qi1454,qi1455,qbc35,qi1473} carefully, this is not surprising
because of the following two reasons.

(1) These proofs are valid for two ensembles with exactly the same density
matrix that spans the same subspace. In contrast, in our case, the two
density matrices corresponding to the sets $\{\left\vert \phi
_{i+}\right\rangle \}$ and $\{\left\vert \phi _{i-}\right\rangle \}$ merely
approach to each other in the $n\rightarrow \infty $\ limit. For any finite $%
n$, the subspaces that they span are never exactly the same. The proofs of
the HJW theorem were never claimed to be applied to such cases.

(2) Those proofs merely predicted the existence of two measurements $M_{\psi
}$ and $M_{\psi ^{\prime }}$. They have not shown explicitly how different
the two measurements are. Our above analysis clearly gives the relationship
between the two measurements $M_{+}$ and $M_{-}$ on $\alpha $ for collapsing
$\beta $ to an element of $\{\left\vert \phi _{i+}\right\rangle \}$ and $%
\{\left\vert \phi _{i-}\right\rangle \}$, respectively. That is, the
corresponding measurement bases are $\{\left\vert \alpha _{i+}\right\rangle
,i=0,...,n-1\}$\ and $\{\left\vert \alpha _{i-}\right\rangle =-\left\vert
\alpha _{i+}\right\rangle ,i=0,...,n-1\}$), where the difference is merely a
global negative sign which has no physical meaning. Thus we can see that
although our final outcome (that steering between these two specific sets is
impossible) looks different from the prediction of the HJW theorem (as
summarized above), there is no theoretical conflict between the proofs of
that theorem and ours.

The properties of $\{\left\vert \phi _{i+}\right\rangle \}$ and $%
\{\left\vert \phi _{i-}\right\rangle \}$ that we found above could make them
very useful in quantum cryptography. There are many no-go theorems in this
field, such as the impossibility of unconditionally secure quantum bit
commitment \cite{qi24,qi23}, and the insecurity proof of two-party quantum
secure computations \cite{qi149}. They are all based directly on the
above-mentioned conclusion of the HJW theorem. In brief, the cheater begins
these quantum cryptographic tasks using entangled states, although in the
honest case he is expected to use either $\psi _{i}$ or $\psi _{i}^{\prime }$
alone, where the density matrices of the sets $\{\psi _{i}\}$ and $\{\psi
_{i}^{\prime }\}$\ are required to be identical (so that the other party
cannot distinguish them). At the end of the process, he chooses between the
two measurements $M_{\psi }$ and $M_{\psi ^{\prime }}$\ on his own system $%
\alpha $, so that system $\beta $ held by the other party will collapse
either to $\psi _{i}$ or $\psi _{i}^{\prime }$ at the cheater's own choice,
just as it is described in the HJW theorem. With this method, the cheater
can gain extra advantage than what is allowed in the honest case, making the
corresponding cryptographic protocol insecure. But our result shows that
when $\psi _{i}$ and $\psi _{i}^{\prime }$ are taken as $\left\vert \phi
_{i+}\right\rangle $ and $\left\vert \phi _{i-}\right\rangle $ in Eqs. (\ref%
{set+}) and (\ref{set-}), the two measurements $M_{\psi }$ and $M_{\psi
^{\prime }}$\ become indistinguishable for infinite-dimensional systems, so
that the cheater no longer has the freedom to choose the state to which $%
\beta $ will collapse. Meanwhile, the two sets $\{\left\vert \phi
_{i+}\right\rangle \}$ and $\{\left\vert \phi _{i-}\right\rangle \}$ also
meet the requirement that they can be indistinguishable to the other party.
Therefore, quantum protocols built upon the states $\{\left\vert \phi
_{i+}\right\rangle \}$ and $\{\left\vert \phi _{i-}\right\rangle \}$ may
eventually evade the cheating strategy in these no-go theorems. We will
study such protocols in forthcoming works.

\section{Summary}

We showed that when a finite-dimensional bipartite system $\alpha \otimes
\beta $\ is prepared in the state described in Eq. (\ref{omega+}), it can
display a chaos effect in quantum steering, such that a subtle difference in
the measurement results on $\alpha $\ can lead to completely orthogonal
steering results on $\beta $.

For infinite-dimensional systems, we showed that it is impossible to
construct a bipartite system $\alpha \otimes \beta $ which can steer the
state of system $\beta $\ between the two sets defined in Eqs. (\ref{set+})
and (\ref{set-}), even though the trace distance between the density
matrices of the two sets can be made arbitrarily small as $n$ increases.
%approach to be identical when $n\rightarrow \infty $.
This result could be very useful for quantum cryptography.

It is also interesting to study whether there exist other forms of states
which can lead to similar anomalous behavior in steering, especially in
finite-dimensional systems.

\bigskip %\section*{Acknowledgements}

The author thanks Dr. Shuming Cheng for helpful discussions. %The work was
%supported in part by the National Science Foundation of
%Guangdong province.
%China.

\appendix

\section{The trace distance}

%\textit{The density matrix.}
Let $\rho _{+}$\ and $\rho _{-}$ denote the density matrices of the two sets
of evenly distributed states $\{\left\vert \phi _{i+}\right\rangle \}$ and $%
\{\left\vert \phi _{i-}\right\rangle \}$ defined in Eqs. (\ref{set+}) and (%
\ref{set-}), respectively. Here we calculate the trace distance between $%
\rho _{+}$\ and $\rho _{-}$.

Denote%
\begin{eqnarray}
\rho _{0\pm i} &\equiv &\frac{1}{\sqrt{2}}(\left\vert 0\right\rangle \pm
\left\vert i\right\rangle )\frac{1}{\sqrt{2}}(\left\langle 0\right\vert \pm
\left\langle i\right\vert )  \nonumber \\
&=&\frac{1}{2}\left[
\begin{array}{c}
1 \\
0 \\
\vdots  \\
0 \\
\pm 1 \\
0 \\
\vdots  \\
0%
\end{array}%
\right] \left[
\begin{array}{cccccccc}
1 & 0 & \cdots  & 0 & \pm 1 & 0 & \cdots  & 0%
\end{array}%
\right] =\left[
\begin{array}{cccccccc}
\frac{1}{2} & 0 & \cdots  & 0 & \pm \frac{1}{2} & 0 & \cdots  & 0 \\
0 & 0 & \cdots  & 0 & 0 & 0 & \cdots  & 0 \\
\vdots  & \vdots  & \ddots  & \vdots  & \vdots  & \vdots  & \vdots  & \vdots
\\
0 & 0 & \cdots  & 0 & 0 & 0 & \cdots  & 0 \\
\pm \frac{1}{2} & 0 & \cdots  & 0 & \frac{1}{2} & 0 & \cdots  & 0 \\
0 & 0 & \cdots  & 0 & 0 & 0 & \cdots  & 0 \\
\vdots  & \vdots  & \vdots  & \vdots  & \vdots  & \vdots  & \ddots  & \vdots
\\
0 & 0 & \cdots  & 0 & 0 & 0 & \cdots  & 0%
\end{array}%
\right] .  \nonumber \\
&&
\end{eqnarray}%
Then for any finite $n$, when each state $(\left\vert 0\right\rangle \pm
\left\vert i\right\rangle )/\sqrt{2}$\ ($i=1,2,...,n-1$) in Eq. (\ref{set+})
and (\ref{set-}) occurs with equal probabilities $1/(n-1)$, the
corresponding density matrices are%
\begin{eqnarray}
\rho _{\pm } &=&\frac{1}{n-1}\sum\limits_{i=1}^{n-1}\rho _{0\pm i}=\left[
\begin{array}{ccccccc}
\frac{1}{2} & \pm \frac{1}{2(n-1)} & \pm \frac{1}{2(n-1)} & \pm \frac{1}{%
2(n-1)} & \cdots  & \pm \frac{1}{2(n-1)} & \pm \frac{1}{2(n-1)} \\
\pm \frac{1}{2(n-1)} & \frac{1}{2(n-1)} & 0 & 0 & \cdots  & 0 & 0 \\
\pm \frac{1}{2(n-1)} & 0 & \frac{1}{2(n-1)} & 0 & \cdots  & 0 & 0 \\
\pm \frac{1}{2(n-1)} & 0 & 0 & \frac{1}{2(n-1)} & \cdots  & 0 & 0 \\
\vdots  & \vdots  & \vdots  & \vdots  & \ddots  & \vdots  & \vdots  \\
\pm \frac{1}{2(n-1)} & 0 & 0 & 0 & \cdots  & \frac{1}{2(n-1)} & 0 \\
\pm \frac{1}{2(n-1)} & 0 & 0 & 0 & \cdots  & 0 & \frac{1}{2(n-1)}%
\end{array}%
\right] .  \nonumber \\
&&
\end{eqnarray}%
Therefore,%
\begin{equation}
\rho _{+}-\rho _{-}=\left[
\begin{array}{ccccc}
0 & \frac{1}{n-1} & \frac{1}{n-1} & \cdots  & \frac{1}{n-1} \\
\frac{1}{n-1} & 0 & 0 & \cdots  & 0 \\
\frac{1}{n-1} & 0 & 0 & \cdots  & 0 \\
\vdots  & \vdots  & \vdots  & \ddots  & \vdots  \\
\frac{1}{n-1} & 0 & 0 & \cdots  & 0%
\end{array}%
\right] ,
\end{equation}%
and we have%
\begin{equation}
(\rho _{+}-\rho _{-})^{^{\dag }}(\rho _{+}-\rho _{-})=\left[
\begin{array}{ccccc}
\frac{1}{n-1} & 0 & 0 & \cdots  & 0 \\
0 & \frac{1}{(n-1)^{2}} & \frac{1}{(n-1)^{2}} & \cdots  & \frac{1}{(n-1)^{2}}
\\
0 & \frac{1}{(n-1)^{2}} & \frac{1}{(n-1)^{2}} & \cdots  & \frac{1}{(n-1)^{2}}
\\
\vdots  & \vdots  & \vdots  & \ddots  & \vdots  \\
0 & \frac{1}{(n-1)^{2}} & \frac{1}{(n-1)^{2}} & \cdots  & \frac{1}{(n-1)^{2}}%
\end{array}%
\right] ,
\end{equation}%
\begin{equation}
\sqrt{(\rho _{+}-\rho _{-})^{\dag }(\rho _{+}-\rho _{-})}=\left[
\begin{array}{ccccc}
\frac{1}{(n-1)^{1/2}} & 0 & 0 & \cdots  & 0 \\
0 & \frac{1}{(n-1)^{3/2}} & \frac{1}{(n-1)^{3/2}} & \cdots  & \frac{1}{%
(n-1)^{3/2}} \\
0 & \frac{1}{(n-1)^{3/2}} & \frac{1}{(n-1)^{3/2}} & \cdots  & \frac{1}{%
(n-1)^{3/2}} \\
\vdots  & \vdots  & \vdots  & \ddots  & \vdots  \\
0 & \frac{1}{(n-1)^{3/2}} & \frac{1}{(n-1)^{3/2}} & \cdots  & \frac{1}{%
(n-1)^{3/2}}%
\end{array}%
\right] ,
\end{equation}%
where $(\rho _{+}-\rho _{-})^{\dag }$ denotes the Hermitian conjugation of $%
(\rho _{+}-\rho _{-})$. Thus the trace distance between $\rho _{+}$ and $%
\rho _{-}$ is%
\begin{equation}
D(\rho _{+},\rho _{-})\equiv \frac{1}{2}tr\sqrt{(\rho _{+}-\rho _{-})^{\dag
}(\rho _{+}-\rho _{-})}=\frac{1}{\sqrt{n-1}}.
\end{equation}

\end{document}